\documentclass[fleqn,10pt]{SelfArx} % Document font size and equations flushed left

\definecolor{color1}{RGB}{0,0,90} % Color of the article title and sections
\definecolor{color2}{RGB}{0,20,20} % Color of the boxes behind the abstract and headings

\usepackage{hyperref} % Required for hyperlinks
\hypersetup{hidelinks,colorlinks,breaklinks=true,urlcolor=color2,citecolor=color1,linkcolor=color1,bookmarksopen=false,pdftitle={Title},pdfauthor={Author},urlcolor=blue}

\usepackage{graphicx}
\usepackage[square]{natbib}
\usepackage{amsmath}
\usepackage{multirow}
\usepackage{subfigure}
\usepackage{pdflscape}

\newcommand{\inte}{\textrm{\scriptsize{in}}}
\newcommand{\exte}{\textrm{\scriptsize{ex}}}
\newcommand{\n}[1]{\mathrm{#1}}

\JournalInfo{Published in Journal of Magnetism and Magnetic Materials, Vol. 322 (22), 3664-3671, 2010} % Journal information
\Archive{\href{http://dx.doi.org/10.1016/j.jmmm.2010.07.022}{DOI: 10.1016/j.jmmm.2010.07.022}} % Additional notes (e.g. copyright, DOI, review/research article)

\PaperTitle{Comparison of adjustable permanent magnetic field sources} % Article title

\Authors{R. Bj\o{}rk, C. R. H. Bahl, A. Smith and N. Pryds} % Authors
\affiliation{\textit{Department of Energy Conversion and Storage, Technical University of Denmark - DTU, Frederiksborgvej 399, DK-4000 Roskilde, Denmark}} % Author affiliation
\affiliation{*\textbf{Corresponding author}: rabj@dtu.dk} % Corresponding author

 % Keywords - if you don't want any simply remove all the text between the curly brackets
\newcommand{\keywordname}{Keywords} % Defines the keywords heading name

\Abstract{A permanent magnet assembly in which the flux density can be altered by a mechanical operation is often significantly smaller than comparable electromagnets and also requires no electrical power to operate. In this paper five permanent magnet designs in which the magnetic flux density can be altered are analyzed using numerical simulations, and compared based on the generated magnetic flux density in a sample volume and the amount of magnet material used. The designs are the concentric Halbach cylinder, the two half Halbach cylinders, the two linear Halbach arrays and the four and six rod mangle. The concentric Halbach cylinder design is found to be the best performing design, i.e. the design that provides the most magnetic flux density using the least amount of magnet material. A concentric Halbach cylinder has been constructed and the magnetic flux density, the homogeneity and the direction of the magnetic field are measured and compared with numerical simulation and a good agreement is found.}

\begin{document}

\flushbottom % Makes all text pages the same height

\maketitle % Print the title and abstract box

%\tableofcontents % Print the contents section

\thispagestyle{empty} % Removes page numbering from the first page

\section{Introduction}\label{Sec.Introduction}
A homogeneous magnetic field for which the flux density can be controlled is typically produced by an electromagnet. To generate a magnetic flux density of 1.0 T over a reasonably sized gap an electromagnet requires a large amount of power, typically more than a thousand watts, and additionally a chiller is needed to keep the electromagnet from overheating. This makes any application using such an electromagnet very power consuming.

Instead of using an electromagnet a permanent magnet configuration for which the flux density can be controlled by a mechanical operation can be used. A number of such variable permanent magnetic flux sources have previously been investigated separately \cite{Cugat_1994, Mhiochain_1999}, and presented in a brief overview \cite{Coey_2002} but no detailed investigations determining the relative efficiencies of the different designs have been published. Here five such designs are compared and the best performing design is found. The efficiency of some of the magnet designs discussed in this paper have also been analyzed elsewhere \cite{Abele_1993,Coey_2003}. However, there only the efficiency of designs of infinite length is characterized. In this paper we consider designs of finite length, which is important as the flux density generated by a finite length magnet assembly is significantly reduced compared to designs of infinite length. Also we parameterize the optimal designs, allowing other researchers to build efficient magnet assemblies.

Examples of applications where an adjustable permanent magnet assembly can be used are nuclear magnetic resonance (NMR) apparatus \cite{Appelt_2006}, magnetic cooling devices \cite{Tura_2007} and particle accelerators \cite{Sullivan_1998}. The flux density source designed in this paper is dimensioned for a new differential scanning calorimeter (DSC) operating under magnetic field designed and built at Ris\o{} DTU \cite{Jeppesen_2008}, but the general results apply for any application in which a variable magnetic field source is needed.

\section{Variable magnetic field sources}
\subsection{Design requirements}
In the analysis of a variable magnetic field source some design constrains must be imposed, such as the minimum and maximum producible flux density. In this analysis the maximum flux density is chosen to be 1.5 T which is a useful flux density for a range of experiments. The minimum flux density is required to be less than 0.1 T both to allow measurements at low values of the magnetic flux density, as well as to allow placement of a sample with only small interaction with the magnetic field. Also a flux density of less than 0.1 T is more easily realizable in actual magnet assemblies than if exactly 0 T had been required.  Ideally the flux density must be homogeneous across the sample at any value between the high and low values. The mechanical force needed to adjust the flux density is also considered.

The magnet assembly must be able to contain a sample that can be exposed to the magnetic field, and the sample must of course be able to be moved in and out of the magnet assembly. The size of a sample can be chosen arbitrarily, and for this investigation a sample volume shaped as a cylinder with a radius of 10 mm and a length of 10 mm was chosen. To allow the sample to be moved we require that the clearance between the magnet and the sample must be at least 2.5 mm, in effect increasing the gap radius to 12.5 mm. The sample volume is sufficiently large to allow the magnet designs to be used in the DSC device discussed above.

\subsection{Numerical analysis}
Given the above design requirements five different permanent magnet designs have been selected for detailed investigation. In each of the designs it is possible to adjust the generated flux density by a mechanical operation. Numerical simulations of each design for a range of parameters were performed and the designs are evaluated based on the mean flux density in the sample volume. Each design was always centered on the sample cylinder.

All numerical work in this paper was done in three dimensions using the commercially available finite element multiphysics program, \emph{Comsol Multiphysics}\cite{Comsol}. The equation solved in the simulations is the magnetic scalar potential equation,
\begin{eqnarray}
-\nabla{}\cdot{}(\mu{}_{0}\mu{}_{r}\nabla{}V_\mathrm{m}-\mathbf{B}_{\mathrm{rem}})=0,\label{Eq.Numerical_equation}
\end{eqnarray}
where $V_\mathrm{m}$ is the magnetic scalar potential, $\mathbf{B}_{\mathrm{rem}}$ is the remanent flux density, $\mu{}_{0}$ is the permeability of free space and $\mu{}_{r}$ is the relative permeability, defined as $\frac{\partial B}{\partial H}$ to account for the remanence of the permanent magnets, and assumed to be isotropic.

Once the magnetic scalar potential has been found, the magnetic field, $\mathbf{H}$, can be found as
\begin{eqnarray}
\mathbf{H} = -\nabla{}V_\mathrm{m} ~,
\end{eqnarray}
and subsequently the magnetic flux density, $\mathbf{B}$, can be determined.

The permanent magnets are modeled by the relation $\mathbf{B} = \mu_0\mu_r\mathbf{H}+\mathbf{B}_{\mathrm{rem}}$, which is justified because the intrinsic coercivity of a NdFeB magnet, which is used as a permanent magnet in present calculations, can be as high as 3 T \cite{Standard}. The transverse susceptibility of the magnets is ignored, as the anisotropy field has a value of 8 T\cite{Zimmermann_1993}. The remanence of the permanent magnets in all designs considered here is $B_{\mathrm{rem}}=1.2$ T and the relative permeability is $\mu_r = 1.05$, in accordance with values for a standard NdFeB magnet \cite{Standard}.

An important issue to note is that the magnetostatic problem is scale invariant, i.e. if all dimensions are scaled by the same factor the magnetic field in a given point will be the same if this point is scaled as well. This means that quantities such as the average value and the homogeneity of the magnetic field in a scaled volume of space will be the same. Thus the conclusions of this paper apply equally to any sample volume that has the same relative dimensions as the sample volume used here, as long as the magnet designs are scaled appropriately.

In the following subsections the five designs are introduced and analyzed.

\subsection{Concentric Halbach cylinders}
The concentric Halbach cylinder consists of two Halbach cylinders, which are cylindrical permanent magnet assemblies that have a direction of magnetization that changes continuously as, in polar coordinates,
\begin{eqnarray}
B_{\mathrm{rem},r}    &=& B_{\mathrm{rem}}\; \textrm{cos}(\phi) \nonumber\\
B_{\mathrm{rem},\phi} &=& B_{\mathrm{rem}}\; \textrm{sin}(\phi)\;,\label{Eq.Halbach_magnetization}
\end{eqnarray}
where $B_{\mathrm{rem}}$ is the magnitude of the remanent flux density \cite{Mallinson_1973, Halbach_1980}.

For practical applications the Halbach cylinder is constructed from segments, each with a constant direction of magnetization. A Halbach cylinder with eight segments produces 90\% of the flux density of a perfect Halbach cylinder while a configuration with 16 segments obtains 95\% of the flux density \cite{Bjoerk_2008}.

\begin{figure}[!t]
  \centering
  \includegraphics[width=0.82\columnwidth]{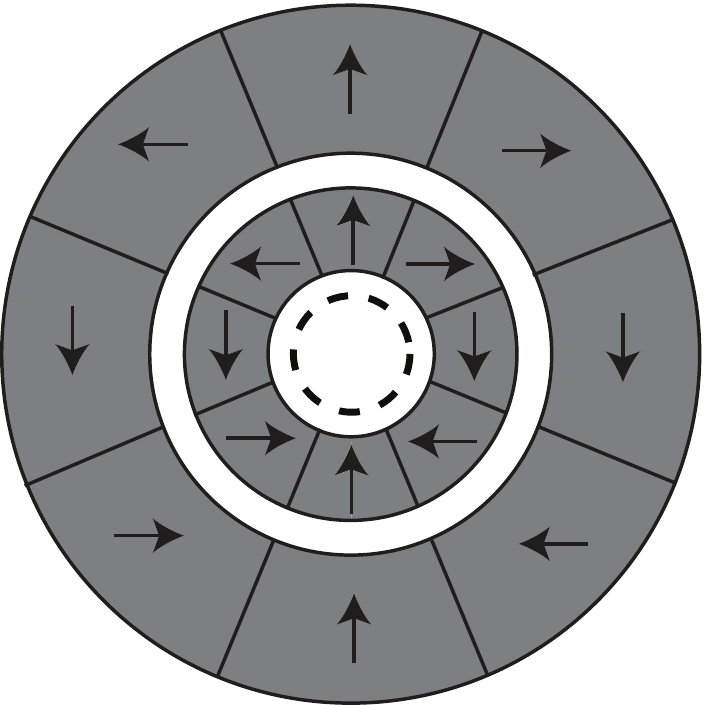}
  \caption{A two dimensional illustration of the concentric Halbach cylinder. Each Halbach cylinder is segmented into eight parts. Shown as arrows is the direction of magnetization. The sample volume is shown as a dashed circle. In the configuration shown the total field in the sample volume is maximized.}
  \label{Fig_Concentric_Halbach_cylinders}
\end{figure}

If two Halbach cylinders are placed concentrically inside each other, the flux density in the inner cylinder bore can be adjusted by rotating one of the cylinders relative to the other. If the permanent magnets used to construct the Halbach cylinders have a permeability close to one, as is the case for NdFeB magnets, the total flux density of the concentric Halbach cylinder is approximately the vector sum of the flux densities produced by the individual cylinders. An illustration of the concentric Halbach cylinder design is shown in Fig. \ref{Fig_Concentric_Halbach_cylinders}.

The concentric Halbach cylinder system is characterized by eight parameters, namely the internal radius, $r_{\inte}$, external radius, $r_{\exte}$, and the length, $L$, of each of the two cylinders, and the number of segments of each cylinder. The segments of the two cylinders were always aligned in the high field position, as shown in Fig. \ref{Fig_Concentric_Halbach_cylinders}.

The advantages of the concentric Halbach cylinder design is that adjusting the flux density by rotating either of the cylinders does not change the geometry of the device. Also, in the infinite length case with no segmentation, there is no torque when rotating one of the cylinders \cite{Bjoerk_2010}.  However, a small torque is present in real-world assemblies, due to segmentation and flux leakage through the cylinder bore \cite{Mhiochain_1999}. The disadvantage of the concentric Halbach cylinders is that even though the cylinders are designed to have exactly the same flux density in the center of the cylinder bore, so that the flux density will be zero when they are offset by $180$ degree, this will not completely cancel the magnetic field away from the center of the bore. This is because the cylinders have different internal radii which means that the flux loss through the ends of each cylinder will not be the same and the flux density will not cancel all the way out of the cylinder bore. This can be important when placing samples in the magnet, as they will respond to the gradient of the magnetic field as they are moved in and out of the cylinder bore.

The parameters varied for the modeling of this design are presented in Table \ref{Table.Varied_parameters}. The internal radius of the outer Halbach cylinder was kept fixed at the external radius of the inner Halbach cylinder plus 2 mm to allow room for the inner cylinder to rotate. Both the inner and outer Halbach cylinder were modeled from eight segments to make the design economically affordable. Many of the above configurations do not produce a sufficiently low magnetic flux density ($< 0.1$ T) in the bore when the cylinders are oppositely aligned. These are not suitable designs and were not considered further.

The field in the exact center of a finite length Halbach cylinder can be calculated analytically by the expression\cite{Zijlstra_1985}
\begin{eqnarray}\label{Eq.Finite_length_Halbach}
B(r=0, z=0) = B_\n{rem}\left(\n{ln}\left(\frac{r_\exte}{r_\inte}\right)+\frac{z_0}{2\sqrt{z_0^2+r_\inte^2}}\right.\nonumber\\
\left.-\frac{z_0}{2\sqrt{z_0^2+r_\exte^2}}-\n{ln}\left(\frac{z_0+\sqrt{z_0^2+r_\exte^2}}{z_0+\sqrt{z_0^2+r_\inte^2}}\right)\right)
\end{eqnarray}
where $z_0=L/2$. The calculated flux density must be corrected for segmentation of the Halbach cylinder. Using this expression the parameters that do not produce a concentric Halbach cylinder for which the field in the center is zero, could also have been found and disregarded. This expression is later compared with the results of the numerical simulations.

%For the concentric Halbach cylinder design, which is described by eight parameters, four parameters were varied and four kept constant. The latter were the internal radius of the inner Halbach cylinder which was kept constant at 12.5 mm to accommodate the sample and the internal radius of the outer Halbach cylinder which was kept fixed at the external radius of the inner Halbach cylinder plus 2 mm to allow room for the inner cylinder to rotate. Also both the inner and outer Halbach cylinder were constructed from eight segments. This is chosen to make the design economically affordable. The first of the four varied parameters was the external radius of the inner Halbach cylinder which was varied from 21-37 mm in 8 nonequidistant steps. The external radius of the outer Halbach cylinder was varied in 8 nonequidistant steps, based on the value of the external radius of the inner Halbach cylinder. The overall minimum value of the external radius of the outer Halbach cylinder was 37 mm, while the maximum was 115 mm. Finally the lengths of the two cylinders were varied individually from 35-95 mm in steps of 10 mm. In total 3136 simulations were performed. Many of the above configurations do not produce a substantially low magnetic flux density ($< 0.1$ T) in the bore when the cylinders are oppositely aligned. These are not suitable designs and were removed, decreasing the number of useable parameterized concentric Halbach cylinders to 709.

\begin{table*}[!t]
\begin{center}
\caption{The parameters varied of each design. The number in parentheses denotes the step size. A asterisk denotes non-equidistant steps and no parentheses indicates a fixed value. For the two linear Halbach array $^{\textrm{a}}$ denotes the width and $^{\textrm{b}}$ denotes the height of a magnet block.}\label{Table.Varied_parameters}
\begin {tabular}{l|rrrrrr}
 & \multicolumn{2}{c}{Concentric} & Two half  & Two linear & Four   & Six \\
 & \multicolumn{2}{c}{Halbach}    & Halbach   & Halbach   & rod    & rod \\
 & inner magnet & outer magnet    & cylinders & array      & mangle & mangle \\
 \hline

Inner radius [mm] & 12.5       & 21-37 (*) + 2                  & 12.5         & 25-150$^{\textrm{a}}$ (5)  & -             & - \\
Outer radius [mm] & 21-37 (*)  & 37-115 (*)                     & 30-150 (10)  & 25-150$^{\textrm{b}}$ (5)  & 10-100 (2.5)  & 1-70 (1) \\
Length       [mm] & 35-95 (10) & 35-95 (10)                     & 30-300 (10)  & 25-150 (5)                 & 10-250 (5)    & 10-600 (5) \\
Segments/rods     & 8          & 8                              & 10           & 3                          & 4             & 6 \\ \hline
\end {tabular}
\end{center}
\end{table*}

% Instead of varying the outer radius of the inner Halbach cylinder directly the predicted two dimensional flux density of the inner Halbach cylinder was varied instead. For a two dimensional Halbach cylinder the flux density is given \cite{Halbach_1980}
% \begin{eqnarray}\label{Eq.Halbach.analytic}
% B = B_{\mathrm{r}}\textrm{ln}\left(\frac{r_{\exte}}{r_{\inte}}\right),
% \end{eqnarray}
% where $B_{\mathrm{r}}$ is the remanent flux density of the magnet material and $r_{\exte}$ and $r_{\inte}$ are the internal and external radii respectively.

% Thus by choose $B$ as a parameter the external radius is given as
% \begin{eqnarray}\label{Eq.Halbach.analytic}
% r_{\exte} = r_{\inte} \mathrm{exp}\left(\frac{B}{B_{\mathrm{r}}}\right).
% \end{eqnarray}

% This approach is chosen to better be able to judge the difference in flux density produced by the inner and outer magnet.

% The two dimensional flux density of the inner Halbach cylinder was varied from 0.6 T to 1.3 T in steps of 0.1 T. This corresponds to a change of the external radii from 2.0609 cm to 3.6931 cm in 8 nonequidistant steps.

% For the internal radius of the outer magnet the outer radius of the inner magnet plus 0.2 cm was used. Thus this was varied from 2.2609 cm to 3.8931 cm in 8 nonequidistant steps.

% The external radius of the outer Halbach cylinder was varied in the same way as for the inner cylinder, i.e. the two dimensional flux density was varied from 0.6 T to 1.3 T in steps of 0.1 T. Thus the external radius assumes a minimum value of 3.7276 cm and a maximum of 11.5023 cm.

\subsection{Two half Halbach cylinders}
\begin{figure}[!t]
  \centering
  \includegraphics[width=1\columnwidth]{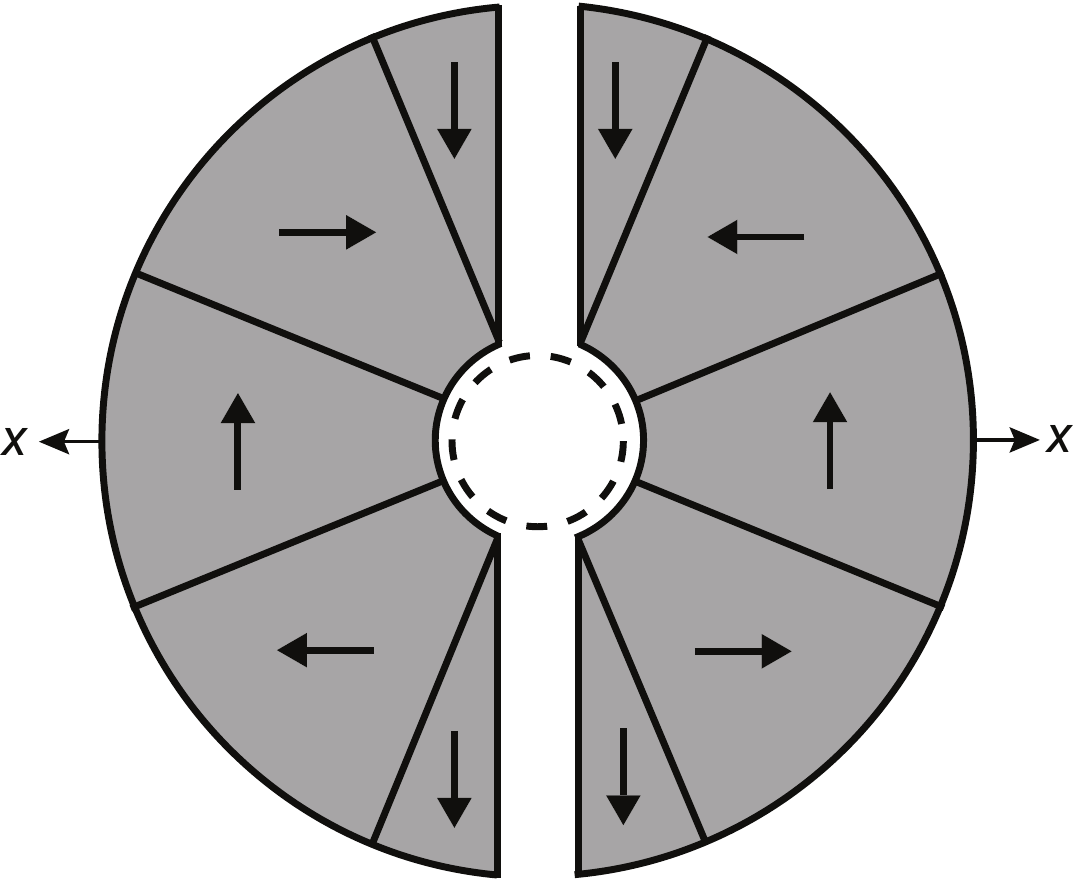}
  \caption{A two dimensional illustration of the two half Halbach cylinders. In total 10 segments are used, of which several are identical. The direction of magnetization is shown as arrows. The sample volume is shown as a dashed circle. Notice the top and bottom gaps between the half-cylinders. This allows room for handling and securing the magnets. The half-cylinders are moved along the $x$-direction to control the flux density.}
  \label{Fig_Halbach_cylinder_10_segments}
\end{figure}

As previously mentioned it is not possible to adjust the flux density of a single particular Halbach cylinder. However, if the Halbach cylinder is split into two parts that can be moved away from each other the flux density between the half-cylinders can be controlled in this way. An illustration of this idea is shown in Fig. \ref{Fig_Halbach_cylinder_10_segments}. This design is termed the two half Halbach cylinders. The design can be characterized by four parameters, namely the internal and external radii and the length of the identical half-cylinders as well as the number of segments. Notice that an additional gap has been included by removing some of the magnet from the top and bottom between the half-cylinders. This has been done to allow room for handling and securing the magnets.

The advantage of this design is that only a simple linear displacement is needed to control the flux density between the cylinders. However, the disadvantage is that there must be enough room to move the half-cylinders away from each other to lower the flux density, and when the half-cylinders are apart the flux density they each generate will influence nearby magnetic objects. Also, a substantial force will in some cases be needed to keep the two half Halbach cylinders close to each other to generate a high flux density.

The parameters varied for this design can again be seen in Table \ref{Table.Varied_parameters}. The number of segments was fixed at ten, again to make the design economically affordable.

%For the two half Halbach cylinders the internal radius was fixed at 12.5 mm and the magnets were designed such that the two half cylinders were always separated by an air gap of 9.5 mm, to make room for handling and securing the magnets, as shown on Fig. \ref{Fig_Halbach_cylinder_10_segments}. The external radius of the two half Halbach cylinders was varied from 30-150 mm in steps of 10 mm and the length of the magnets was varied from 30-300 mm in steps of 10 mm. The total number of segments used to construct the Halbach cylinder was fixed at sixteen.

\subsection{Two linear Halbach arrays}
\begin{figure}[!t]
  \centering
  \includegraphics[width=1\columnwidth]{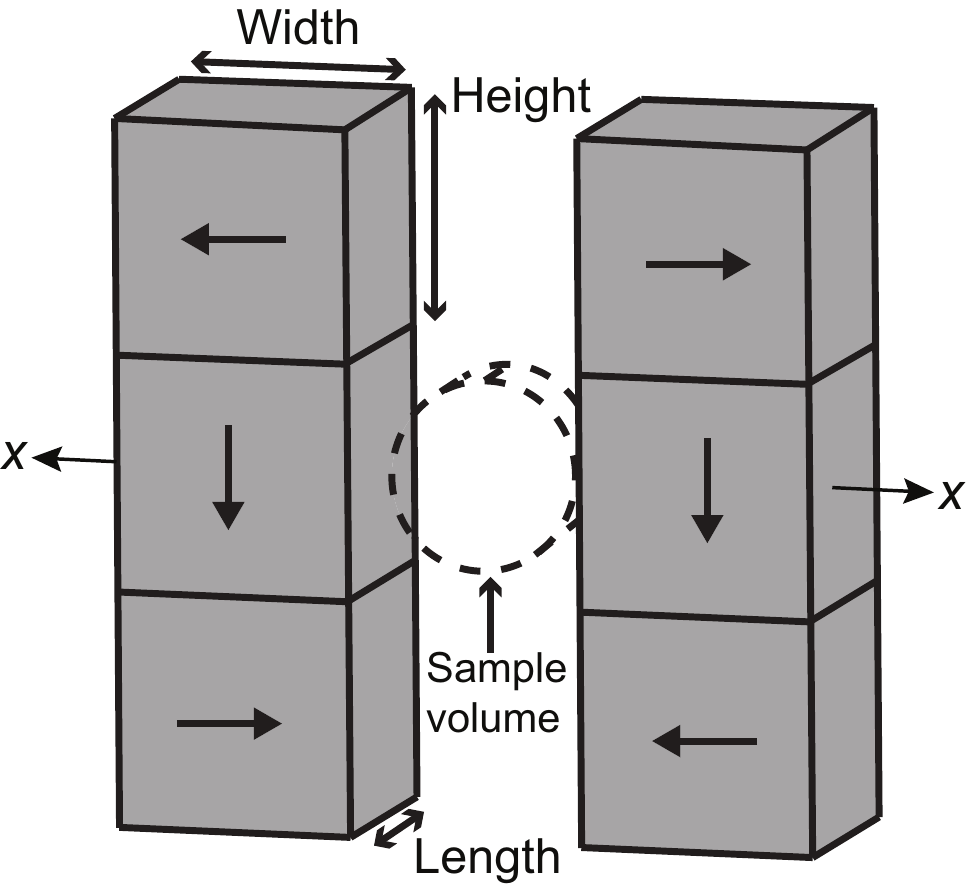}
  \caption{A three dimensional illustration of two three block linear Halbach arrays. The high flux density region is created in between the two arrays, where the sample volume is placed. The arrays are moved along the $x$-direction to control the flux density.}
  \label{Fig_Halbach_array}
\end{figure}

The linear Halbach array is a magnetic assembly that uses the same principle as the Halbach cylinder to generate a one-sided flux \cite{Mallinson_1973}. The linear Halbach array is characterized by the width, height and length of the identical blocks as well as the number of blocks used in the array. For the array considered here three blocks are used, as this is the minimum number of blocks needed to create a one-sided array. An adjustable flux density configuration can be made by placing two mirrored linear Halbach arrays opposite each other, as with the two half Halbach cylinders. By moving the arrays closer or further apart the flux density between them can be controlled. An illustration of the two linear Halbach array design is shown in Fig. \ref{Fig_Halbach_array}.

The sample volume can, because of its short length, be rotated, so that the arrays can be placed closer to each other. This configuration has also been considered, although it might require an alternative method for mounting the sample than for the other designs considered here.

The advantage of the linear Halbach array is that it is easy to construct, as it can be made using simple rectangular magnet blocks. However, the design has the same disadvantages as the two half Halbach cylinders in that a large force will, in some cases, be needed to keep the arrays close together and a high flux density will still be generated when the arrays are moved apart, which could influence nearby magnetic objects.

For the sample position as shown in Fig. \ref{Fig_Halbach_array} the two linear Halbach arrays were separated by a distance of 25 mm, so that the sample volume fitted in between the arrays. For the alternative sample orientations the arrays were separated by a distance of 15 mm, so that the rotated sample fitted between the arrays. For either of the sample positions the height, width and length of a rectangular permanent magnet block were independently varied as given in Table \ref{Table.Varied_parameters}. Each array consists of three identical blocks.

One can envision designs that have a geometrical form ``between'' the two linear Halbach arrays and the two half Halbach cylinders. The performance of these will be comparable to the performance of either of the two linear Halbach arrays or the two half Halbach cylinders.

\subsection{The mangle}
\begin{figure}[!t]
  \centering
  \includegraphics[width=1\columnwidth]{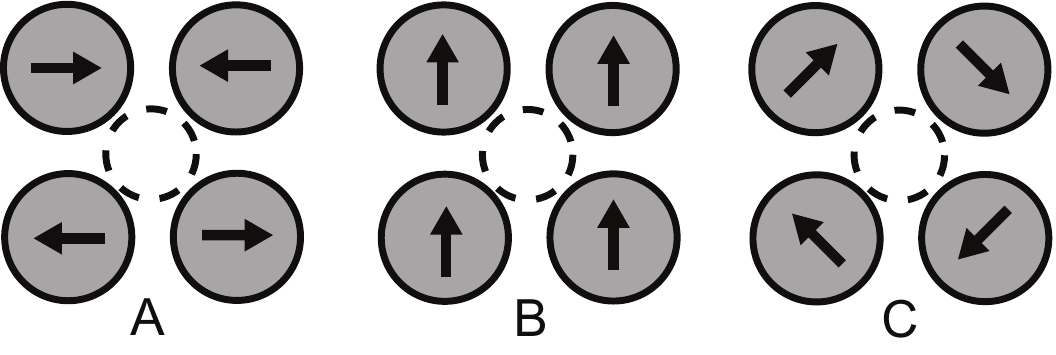}
  \caption{A schematic drawing of the four rod mangle design. (A) shows the high flux density position of the four rod mangle design. The high flux density is created across the sample volume. (B) shows the low flux density configuration, as suggested by \cite{Cugat_1994}. This position is reached by a 90 degree alternate rotation of the rods shown in (A). (C) shows an alternate position of the rods that generate a much lower flux density in the sample volume than the position shown in Fig. B.}
  \label{Fig_Mangle_4}
\end{figure}

The mangle is made up of identical transversely magnetized permanent magnet rods that can be rotated to alter the flux density at the center of the assembly \cite{Cugat_1994} . The rods must be rotated alternately clockwise and counterclockwise to continuously alter the flux density in a homogeneous way. The design can be characterized by three parameters, namely the radius and the length of a rod as well as the number of rods used. An illustration of a mangle design with four cylinders in the orientation that generates a high flux density is shown in Fig. \ref{Fig_Mangle_4} A. The conventional low flux density orientation for the four rod mangle, shown in Fig. \ref{Fig_Mangle_4} B, does not produce a very low flux density, typically around 0.1-0.3 T across the sample volume (if magnet rods with a remanence of 1.2 T are used). An alternative orientation of the rods, shown in Fig. \ref{Fig_Mangle_4} C, produces a much lower flux density, typically less than 0.05 T across the sample volume. Unfortunately there is no way to adjust the flux density from the configuration shown in Fig. \ref{Fig_Mangle_4} A to that shown in Fig. \ref{Fig_Mangle_4} C while maintaining homogeneity in the sample volume. Thus in the four rod mangle considered here we envision a design where the rods are rotated from the configuration shown in Fig. \ref{Fig_Mangle_4} A to that in Fig. \ref{Fig_Mangle_4} B and finally to that in Fig. \ref{Fig_Mangle_4} C.

\begin{figure}[!t]
  \centering
  \includegraphics[width=1\columnwidth]{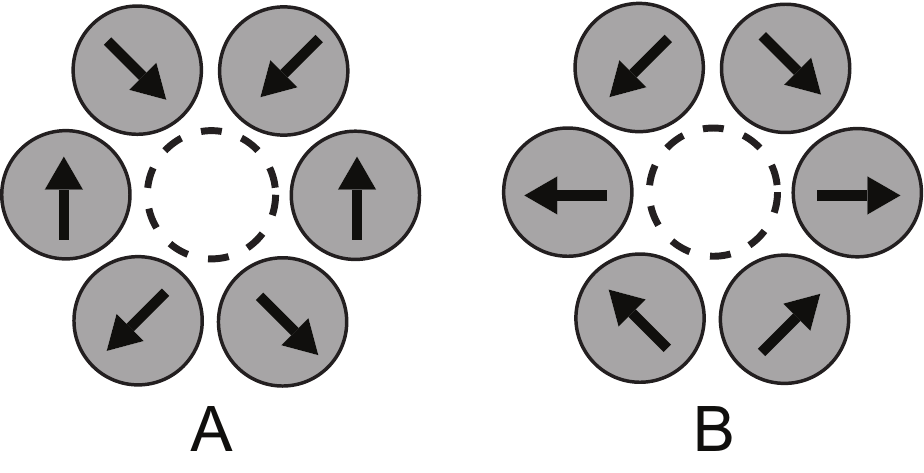}
  \caption{A schematic drawing of the six rod mangle design. The high flux density position of the six rod mangle is shown in Fig. A, while the low flux density position of the six rod mangle in shown in Fig. B. In the latter figure the rods have been alternately rotated 90 degree from the position shown in Fig. A.}
  \label{Fig_Mangle_6}
\end{figure}

A six rod mangle design is also considered. The high flux density orientation of the rods is shown in Fig. \ref{Fig_Mangle_6} A, while the low flux density configuration is shown in Fig. \ref{Fig_Mangle_6} B. Notice that the rods have simply been turned 90 degrees alternately. The low flux density orientation produce a flux density typically less than 0.1 T across the sample volume, so no alternate orientations need be considered.

The advantage of this design is economical as transversely magnetized rods are readily available. The design is also compact and produces a low stray flux density. The disadvantage is that the volume between the rods scales with the radius of the rods. Therefore the flux density can only be increased to a maximum value for a given sample volume, without increasing the size of the volume between the rods.

The parameters varied for both the four and six rod mangles are given in Table \ref{Table.Varied_parameters}. The rods are always placed as closely as possible to each other or to the sample, depending on the mangle parameters.

\section{Comparing the different designs}
To find the best parameters for each design parameter variation simulations were conducted for each of the different designs with the parameters previously stated.

\begin{figure}[!t]
  \centering
  \includegraphics[width=1\columnwidth]{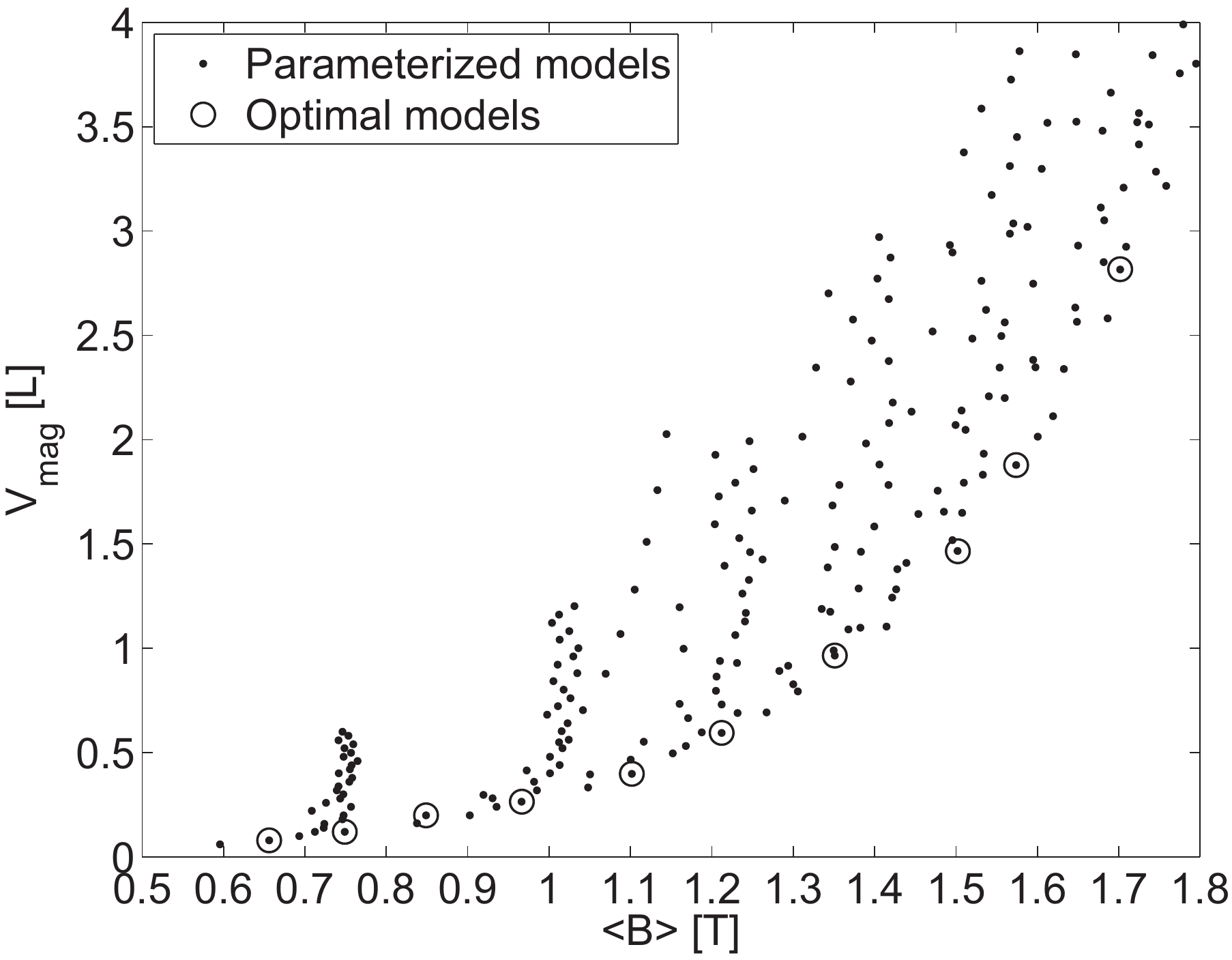}
  \caption{The volume of the magnets as a function of the average flux density in the sample volume for the two half Halbach cylinders design. Some of the optimal designs are marked by circles.}
  \label{Fig_Comp_half_Halbach_cylinders}
\end{figure}

To allow the designs to be more easily compared, the best performing of each of the five different designs are selected. This is done by selecting the parameters that produce a high average flux density in the sample volume and at the same time has a low volume of magnet material. This approach has previously been used to optimize the Halbach cylinder design \cite{Bjoerk_2008}, but other optimization methods exist such as the figure of merit, $M^{*}$, which is almost identical to the optimization used here except that it also include the remanence of the magnets \cite{Coey_2003}. Also an optimization parameter for permanent magnet assemblies used in magnetic refrigeration devices exists \cite{Bjoerk_2010b}. In Fig. \ref{Fig_Comp_half_Halbach_cylinders} both these optimally dimensioned designs, as well as the other parameter variations tried, are shown for the two half Halbach cylinder design. Some of the optimal designs have been indicated in the figure, and it is clearly seen that these produce a given flux density using the lowest amount of magnet material. The same analysis has been performed for the four other designs.

It is worth noting that the designs referred to here as ``optimal'' are not necessarily the global optimal designs. They are the optimal designs of the conducted parameter survey, and as such designs might exist outside the parameter space, or at resolutions smaller than the varied parameters that perform better than the designs referred to as optimal here. However, based on the detail of the parameter survey the potential for improvement will be small.

\subsection{The best parameters for each design}
The parametrization of the optimal designs of each individual design type have been found by analyzing the dimensions of the optimal designs for each flux density. For the concentric Halbach cylinder the optimal designs fulfil
\begin{eqnarray}
0.8 < \frac{r_{\mathrm{inner, ex}}/r_{\mathrm{inner, in}}}{r_{\mathrm{outer, ex}}/r_{\mathrm{outer, in}}} < 1\;\;\mathrm{and}\;\;0.8 < \frac{L_{\mathrm{inner}}}{L_{\mathrm{outer}}} < 1~,
\end{eqnarray}
where the first subscript denotes the inner or outer magnet and the second denotes the internal or external radius. However, this criteria is not enough, as some designs that fulfil these requirements produce a flux density in the low flux density configuration that is higher than 0.1 T. Thus the analytical expression for the flux density in the center of the system, Eq. (\ref{Eq.Finite_length_Halbach}), must be calculated to ensure that this will be less than 0.1 T.

The optimal mangle designs are those where the radius, $r$, of the individual mangle rods is sufficiently small that the rods can be placed close to the sample without touching each other. For the four rod mangle, the relation $r \leq 32.5$ mm applies for the optimal designs. Also, the ratio of the radius to the length must be in the range of $0.2 < \mathrm{radius}/\mathrm{length} < 0.5$. Increasing the length of the rods increases the flux density in the sample volume.

For the six rod mangle only the designs that have $r \leq 10$ mm are optimal. Also, the ratio of the radius to the length must obey $0.05 < \mathrm{radius}/\mathrm{length} < 0.5$, where the lower bond is necessary to obtain a high flux density.

For the linear Halbach array with the sample volume as shown in Fig. \ref{Fig_Halbach_array} the optimal designs are parameterized by
\begin{eqnarray}
\mathrm{height} &\simeq& 0.22\cdot\mathrm{width} + 0.02 \textrm{[mm]}\nonumber\\
\mathrm{length}  &\simeq& 1.0\cdot\mathrm{width} + 0.05 \textrm{[mm]}~,
\end{eqnarray}
where the length, width and height are as shown in Fig. \ref{Fig_Halbach_array}.

For the linear Halbach array with the rotated sample volume the optimal designs are parameterized by
\begin{eqnarray}
\mathrm{height} &\simeq& 0.08\cdot\mathrm{width} + 0.02 \textrm{[mm]}\nonumber\\
\mathrm{length}  &\simeq& 1.3\cdot\mathrm{width} + 0.02 \textrm{[mm]}~,
\end{eqnarray}
where the dimensions are again as shown in Fig. \ref{Fig_Halbach_array}. Because the sample volume has been rotated a much smaller height and a longer length is now favored.

For the two half Halbach cylinders the optimal parameterized designs are characterized by the relation: $\mathrm{radius} \simeq 0.95\cdot\mathrm{length}$. This relation is in agreement with the optimal dimensions for a Halbach cylinder\cite{Bjoerk_2008}.

These parameterizations are obviously only valid for the sample volume with the relative dimensions as chosen here. If a different sample volume were chosen the relations would be different. However, if the sample volume is simply scaled by a factor then, owing to the linearity of the magnetostatic problem, the magnet dimensions need simply be scaled by the same factor to produce the same flux density, and thus in this situation the parameterizations found above remain valid appropriately scaled.

\begin{figure}[!t]
  \centering
  \includegraphics[width=1\columnwidth]{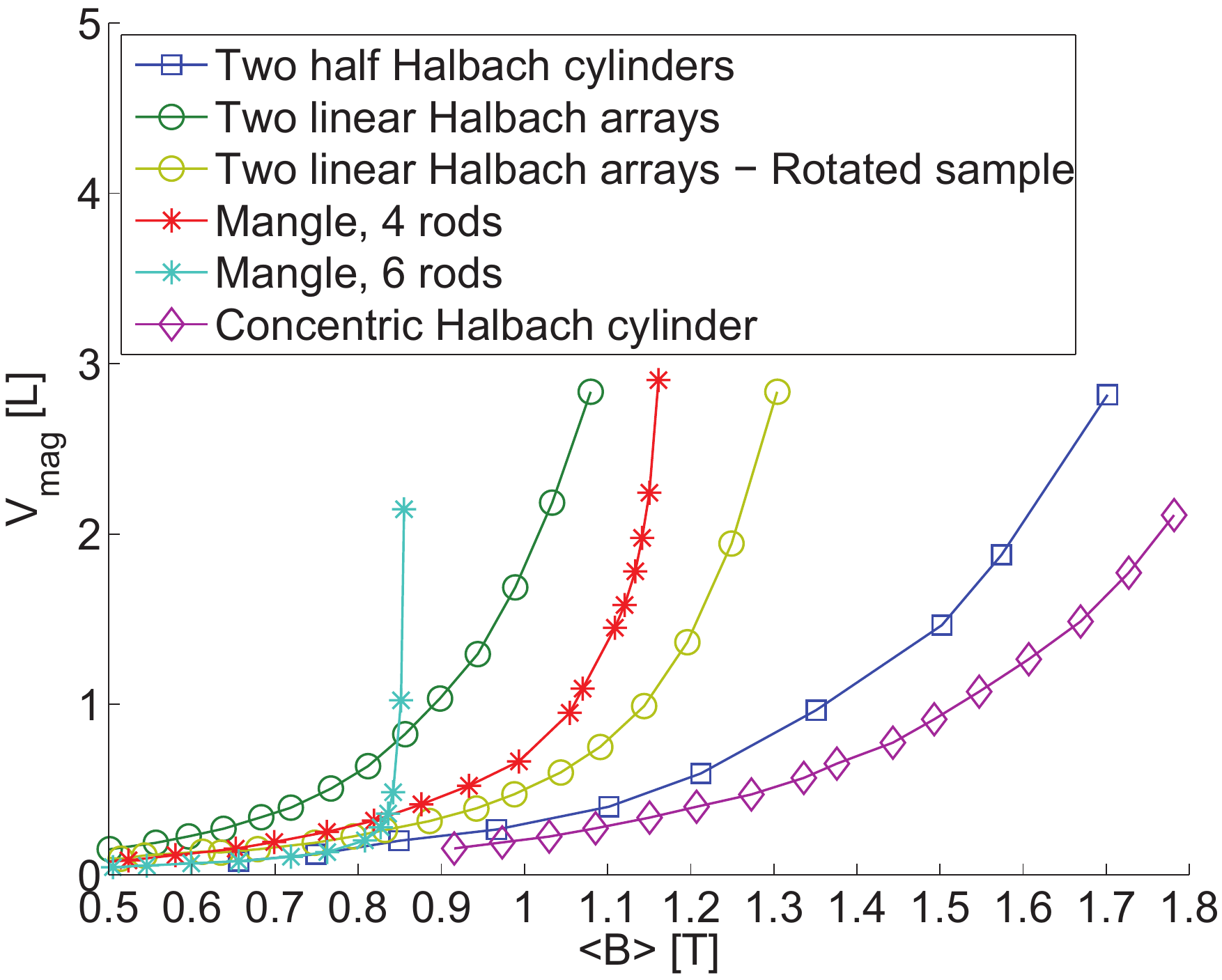}
  \caption{(Color online) The volume of the magnets as a function of the average flux density in the sample volume for the  best individual designs for the six designs considered.}
  \label{Fig_Comp_all}
\end{figure}

The optimal designs for the different design types are shown in Fig. \ref{Fig_Comp_all}. The magnetic flux density produced by a given optimal design, i.e. a design whose dimensions follow the above parameterizations, can be found from Fig. \ref{Fig_Comp_all} by calculating the volume of the magnet in the design.

It is seen that the concentric Halbach cylinder design is the optimal design as it produces a given magnetic flux density using the lowest amount of magnet material. An interesting observation is that the mangle designs are not able to produce a high flux density. This is because, as already stated, as the radius of the rods in the mangle increases the rods must be moved further away from each other, so as not to touch, and thus the volume in between them increases. The two half Halbach cylinders and the concentric Halbach cylinder do not perform identically due to the top and bottom gaps between the half-cylinders and due to the gap between the concentric cylinders.

\begin{figure}[!t]
  \centering
  \includegraphics[width=1\columnwidth]{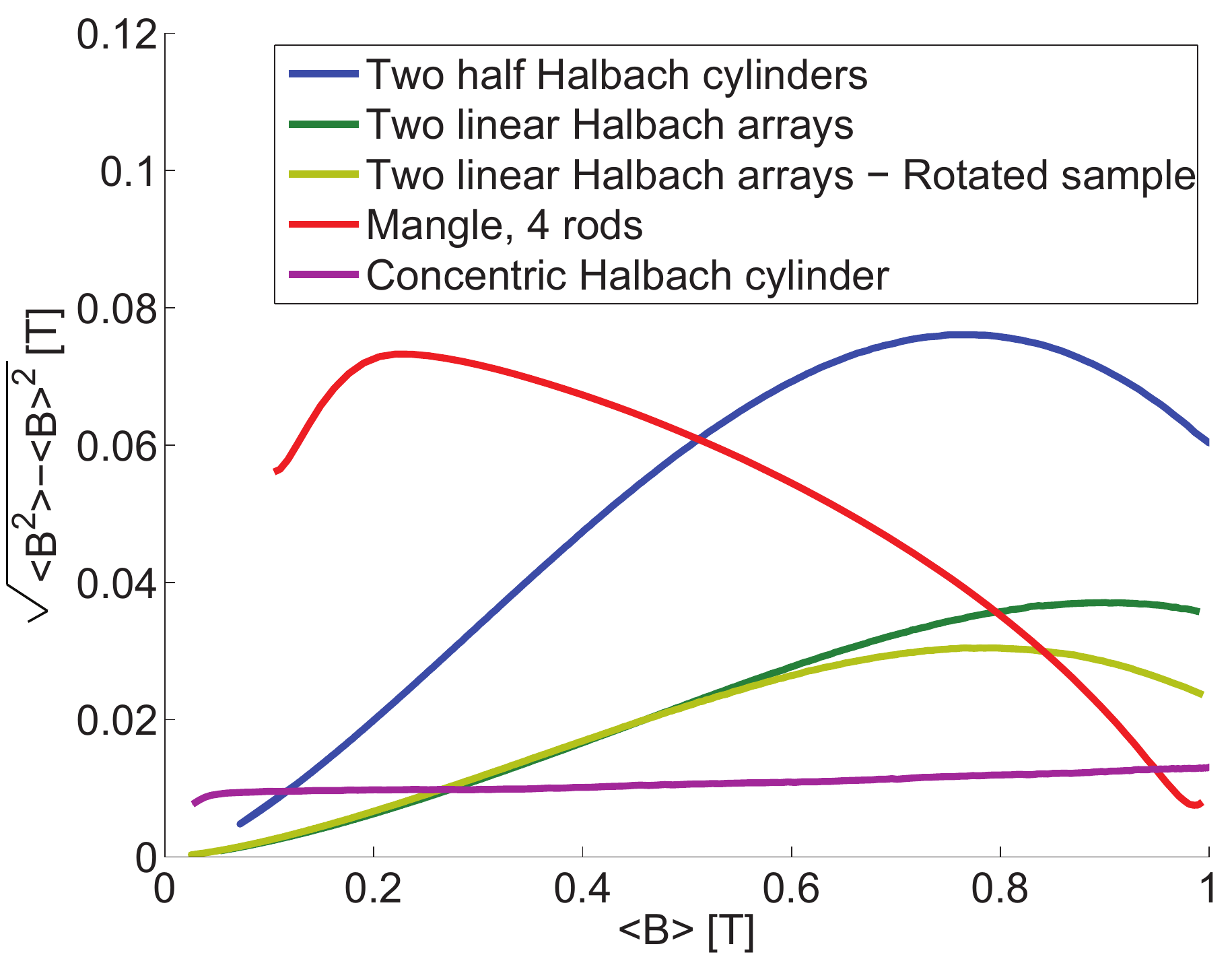}
  \caption{(Color online) The homogeneity, characterized by the standard deviation $\sqrt{\langle{}B^2\rangle{}-\langle{}B\rangle{}^2}$, for the optimal different types of designs that produce 1 T as a function of the average flux density. The mangle was turned from the position shown in Fig. \ref{Fig_Mangle_4} A to that shown in Fig. \ref{Fig_Mangle_4} B.}
  \label{Fig_Homogeneity}
\end{figure}

It is also interesting to consider the homogeneity of the flux density in the sample volume. To characterizes the homogeneity, the best parameter set for each design that produces $1\pm0.01$ T in the high flux density position have been found. The six rod mangle is not able to produce this flux density and so it is not present in the figure. The flux density for these designs have then been varied either by rotation (mangle and concentric Halbach cylinder) or translation (two half Halbach cylinder and two linear Halbach array). Fig. \ref{Fig_Homogeneity} shows the standard deviation of the flux density, $\sqrt{\langle{}B^2\rangle{}-\langle{}B\rangle{}^2}$, as a function of the average flux density, $\langle{}B\rangle{}$, for these optimal 1 T designs. All the design types produce a quite homogeneous flux density across the sample volume, but again the best design is the concentric Halbach cylinder design.

The high homogeneity of the concentric Halbach cylinder means that the difference between the flux density calculated using Eq. (\ref{Eq.Finite_length_Halbach}) and the numerically calculated mean flux density in the sample volume is less than 0.05 T in all considered cases.

\begin{figure}[!t]
  \centering
  \includegraphics[width=1\columnwidth]{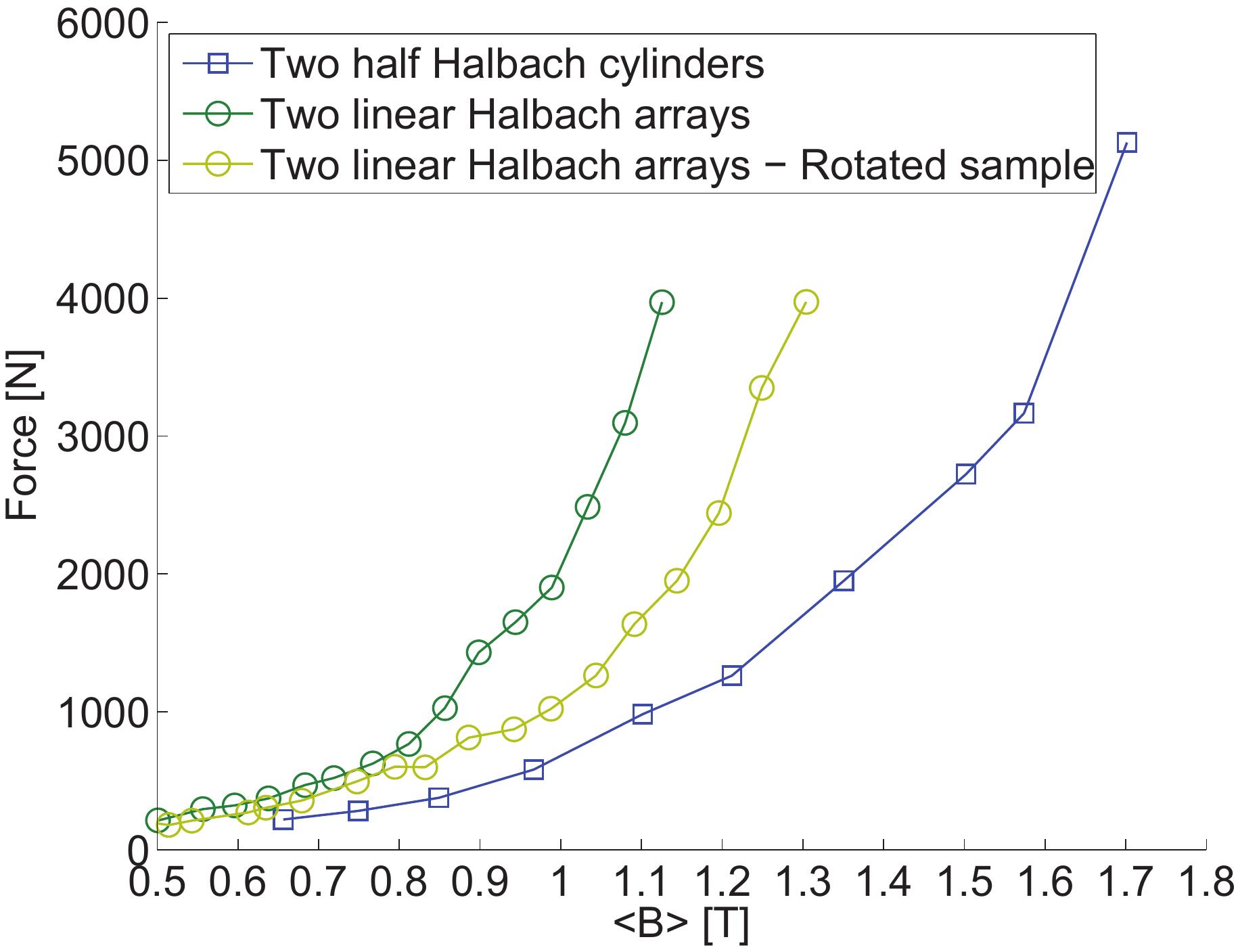}
  \caption{(Color online) The maximum force needed to keep the design at the maximum flux density.}
  \label{Fig_Force}
\end{figure}

In Fig. \ref{Fig_Force} the maximum force as a function of flux density for the two half Halbach cylinder and the linear Halbach array designs are shown. The force shown in the figure is the force on the optimal designs that is needed to keep the two halfs of each design as close together as the sample volume allows. As can be seen a substantial force is needed for the designs that generate a high flux density.

\section{A constructed variable field source}
An adjustable permanent magnet has been built based on the concentric Halbach cylinder design, as this is the best performing and most practical magnet design. A maximum flux density of 1.5 T was chosen as the desired value in the sample volume. The dimensions of the magnet are given in Table \ref{Table.Dimensions}.

\begin{table}[!t]
\begin{center}
\caption{The dimensions of the constructed concentric Halbach cylinder magnet.}\label{Table.Dimensions}
\begin {tabular}{l|cc}
% & Internal radius & External radius & Length \\
% & [mm] & [mm] & [mm] \\ \hline
%Inner magnet   & 12.5  & 27.5 &  55\\
%Outer magnet   & 29.5  & 80   &  65\\
& Inner  & Outer  \\
& magnet & magnet \\ \hline
Internal radius [mm] & 12.5 & 29.5 \\
External radius [mm] & 27.5 & 80 \\
Length [mm]          & 55   & 65
\end {tabular}
\end{center}
\end{table}

\begin{figure}[!t]
  \centering
  \includegraphics[width=1\columnwidth]{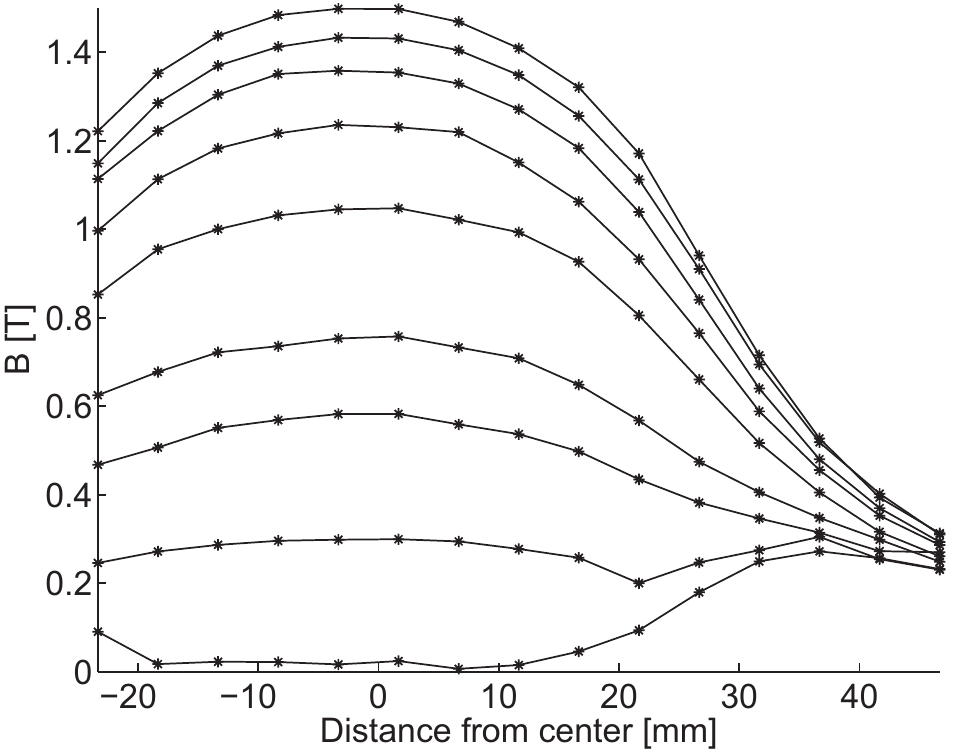}
  \caption{The measured magnetic flux density as a function of distance from the center of the concentric Halbach cylinder. Between each measurement series the cylinders were rotated relative to each other by 22.5 degree. Due to of the design of the magnet it was only possible to measure down to $-25$ mm.}
  \label{Fig_Measured_field_center}
\end{figure}

The magnet was constructed and using a Hall probe (AlphaLab Inc, Model: DCM) the flux density produced by the design was measured. Both components of the magnetic flux density in the plane perpendicular to the cylinder axis, as well as the component parallel to the axis, were measured at 5 mm intervals in the center of the cylinder for nine relative rotation positions of the two cylinders, each separated by 22.5 degrees. The initial angle was chosen to be a close to a flux density of zero as possible. The norm of the vector sum of the three components of the magnetic flux density is shown in Fig. \ref{Fig_Measured_field_center}.

The uncertainty on the position of the Hall probe is estimated to be $\pm 1$ mm. There is also an uncertainty in the 90 degree rotation of the Hall probe necessary to measure the two components of the flux density that are perpendicular to the cylinder axis. It is estimated that these uncertainties result in a total uncertainty of $\pm 5\%$ for the magnetic flux density. The instrumental uncertainty of the Hall probe is $\pm 0.2\%$, which is much less than the uncertainty due to the positioning of the Hall probe. No errorbars are shown in Fig. \ref{Fig_Measured_field_center} in order to maintain clarity in the plot.

The axial component of the magnetic flux density is included in the flux density shown in Fig. \ref{Fig_Measured_field_center}, but is quite small. At no point in the cylinder bore does the axial component exceed 0.15 T for any rotation angle, and in the center it is always less than 0.05 T for any rotation angle.

\begin{figure}[!t]
  \centering
  \includegraphics[width=1\columnwidth]{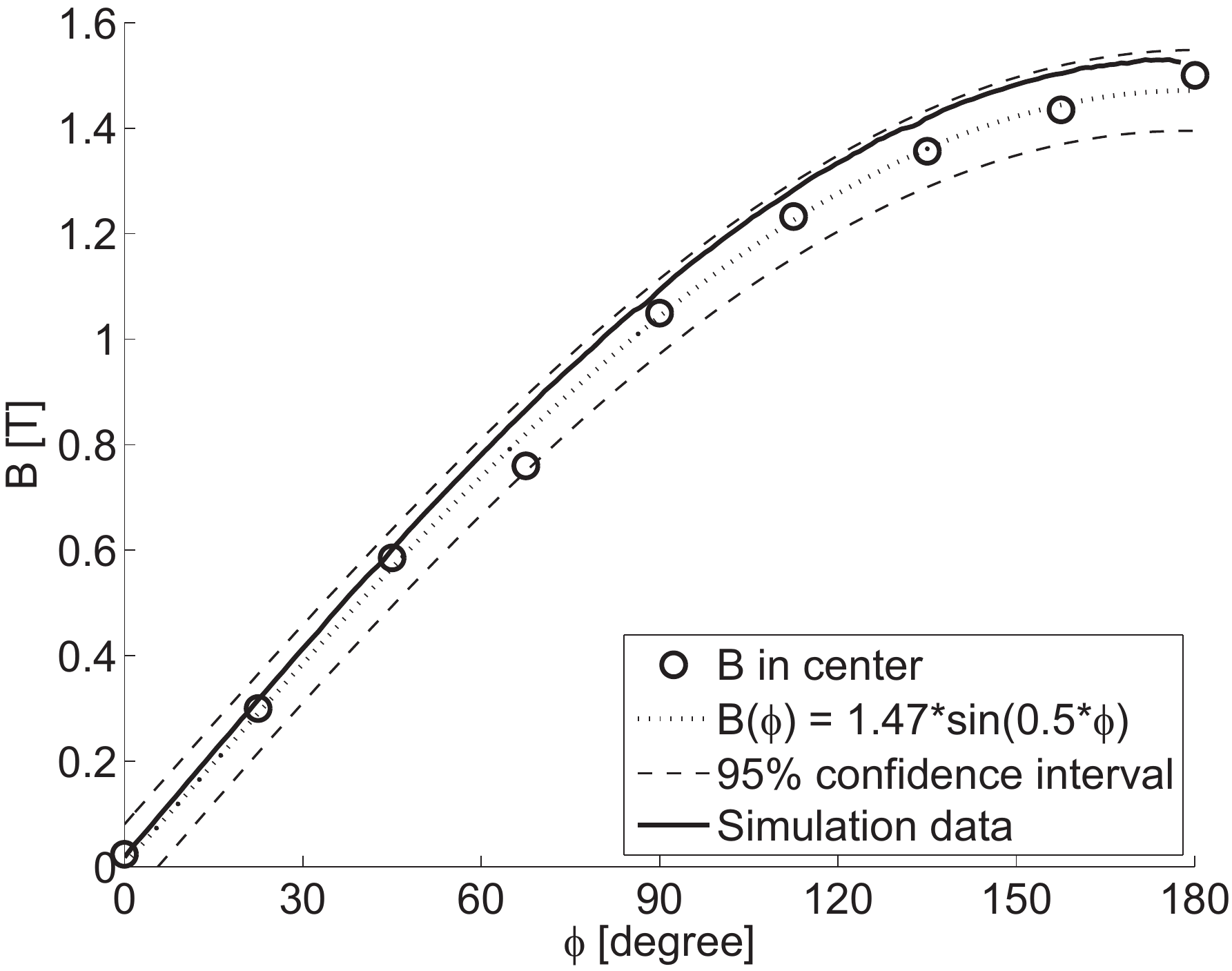}
  \caption{The center value of the flux density as a function of rotation angle, $\phi$.  A sine function has been fitted to the data. Also shown are the results from numerical simulations.}
  \label{Fig_Sinus_fit_rotation}
\end{figure}

The measured values of the magnetic flux density have been interpolated to find the value at the center of the concentric cylinder. These values are shown in Fig. \ref{Fig_Sinus_fit_rotation} as a function of the displacement angle, $\phi$, between the two cylinders. A sine function of the form $B = \alpha\;\mathrm{sin}(0.5(\phi+\beta))$, where $\alpha$ and $\beta$ are constants, has been fitted to the data as this is how the field should theoretically vary. This is so because the magnetic flux density produced by the inner and outer magnet is identical in the center and thus the combined flux density can be found based on law of cosine for an isosceles triangle. The fit is shown in Fig. \ref{Fig_Sinus_fit_rotation} as well as the 95\% confidence interval of the fit for a new measurement. The constants were determined to be $\alpha = 1.47 \pm 0.04$ T and $\beta = 0 \pm 3$ degree.

The magnet design has also been simulated numerically and the resulting flux densities are also shown in Fig. \ref{Fig_Sinus_fit_rotation}. A reasonable agreement between the measured flux density and the value predicted by simulation is seen. It is seen that the flux density can easily be adjusted by rotating the inner cylinder relative to the outer cylinder.

\begin{figure}[!t]
  \centering
  \includegraphics[width=1\columnwidth]{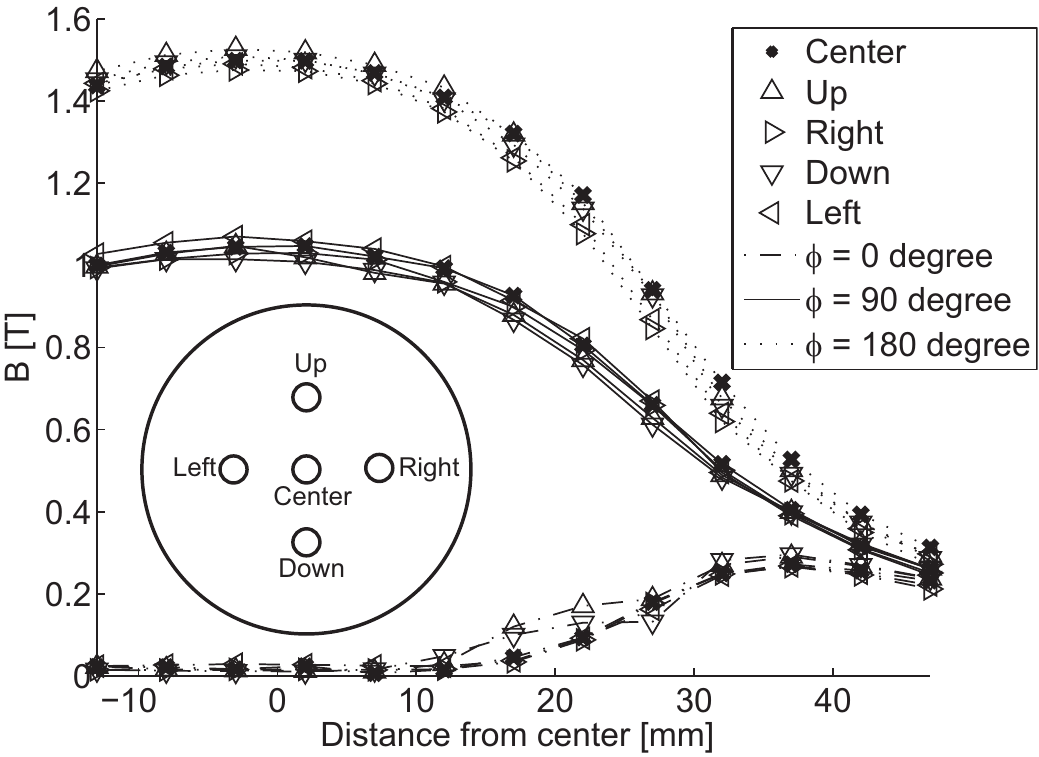}
  \caption{The homogeneity of the measured magnetic flux density as a function of distance from the center of the concentric Halbach cylinder. The positions labeled \emph{Up}, \emph{Right}, \emph{Down} and \emph{Left} are located 5.5 mm from the center along the direction corresponding to, respectively, 0, 90, 180 and 270 degrees.}
  \label{Fig_Homogeneity_measured_field_with_subfigure}
\end{figure}

The homogeneity of the flux density has been investigated by measuring the flux density at four off-center positions. These are located 5.5 mm from the center along an angle corresponding to respectively 0, 90, 180 and 270 degrees. The results for three different displacement angles, $\phi = 0, 90, 180$ degrees respectively, are shown in Fig. \ref{Fig_Homogeneity_measured_field_with_subfigure}. The standard deviation, $\sqrt{\langle{}B^2\rangle{}-\langle{}B\rangle{}^2}$, can be calculated for the sample volume based on the data in Fig. \ref{Fig_Homogeneity_measured_field_with_subfigure}. Using the four central data points to represent the sample volume one obtains $\sqrt{\langle{}B^2\rangle{}-\langle{}B\rangle{}^2} = 0.61\cdot{}10^{-2}, 2.18\cdot{}10^{-2} \textrm{ and } 2.17\cdot{}10^{-2}$ for $\phi = 0, 90 \textrm{ and } 180$ degree respectively. Thus the flux density distribution is quite homogeneous.

\begin{figure}[!t]
  \centering
  \includegraphics[width=1\columnwidth]{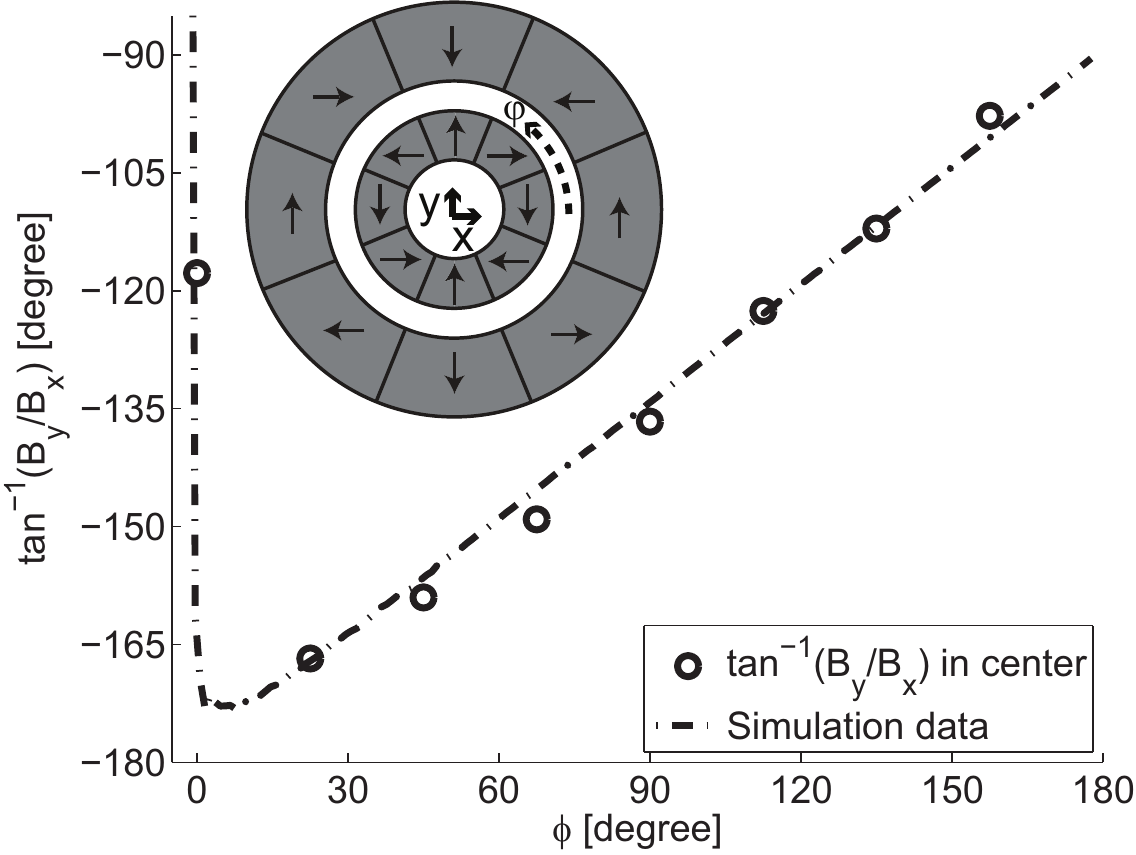}
  \caption{The direction of the field as a function of rotation angle, $\phi$ in the coordinate system shown in the figure. The field changes direction by 180 degree when $\phi = 0$ is crossed, at which point $B = 0$.}
  \label{Fig_Direction_of_field_rotation_with_subfigure}
\end{figure}

The direction of the flux density changes as the cylinders are rotated with respect to each other. Fig. \ref{Fig_Direction_of_field_rotation_with_subfigure} shows the direction as a function of the rotation angle, $\phi$, for the measured flux density as well as simulation data for the coordinate system as shown in the figure. A good agreement between these is seen.

The agreement between the measured magnetic flux density and the simulation results is limited by several factors. A perfect agreement is not expected as the transverse susceptibility for the magnets is ignored. However, the major source of error is estimated to be the positioning and rotation of the Hall probe in the conducted measurements, as described earlier.

For all the five designs it is important to consider the coercivity of the permanent magnets used. For, e.g. the concentric Halbach cylinder design when the cylinders are offset by $180$ degree the flux density produced by the outer cylinder will be parallel and opposite to the remanence of parts of the inner cylinder, and if this flux density is higher than the coercivity of the magnets the direction of magnetization will be reversed, which will render the device useless \cite{Bjoerk_2008,Bloch_1998}. For the simulated permanent magnets a remanence of 1.2 T was used. A typical industry NdFeB magnet with such a remanence has a high coercivity, $\mu_{0}H_{c} = 3.2$ T, which is sufficiently strong to keep the direction of magnetization constant.

\section{Conclusion}
Five different variable permanent magnet designs, the concentric Halbach cylinder, the two half Halbach cylinders, the two linear Halbach arrays and the four and six rod mangles, were investigated and evaluated based on the generated magnetic flux density in a sample volume and the amount of magnet material used. As the dipole field is scale invariant the conclusion holds for all sample volumes with the same relative dimensions as used here. The best performing design, i.e. the design that provides the highest magnetic flux density using the least amount of magnet material, was the concentric Halbach cylinder design. Based on this result a concentric Halbach cylinder was constructed and the magnetic flux density, the homogeneity  and the direction of the magnetic flux density were measured. These were compared with numerical simulation and a good agrement was found.

\section*{Acknowledgements}
The authors would like to thank J. Geyti for his technical assistance. Also, the authors would like to acknowledge the support of the Programme Commission on Energy and Environment (EnMi) (Contract No. 2104-06-0032) which is part of the Danish Council for Strategic Research.

%-------------------------------------------------------------------------------------------------------------
%-------------------------------------------------------------------------------------------------------------
%-------------------------------------------------------------------------------------------------------------

\end{document}